# Half-Metallic $L2_1$ Structures with (001) Planar Insertions

C. A. Culbert, M. Williams[*], M. Chshiev, and W. H. Butler

*Center for Materials for Information Technology, University of Alabama, Tuscaloosa, AL*
[*]*Mathematics and Computer Sciences, University of Maryland Eastern Shore, Princess Anne, MD*

*Abstract*

A number of $L2_1$ phase alloys (composition $X_2YZ$) are half-metallic. Although this structure is typically described in terms of an fcc Bravais lattice with a 4 atom basis, it can be viewed more simply as a variant of bcc or B2 in which planes of $X_2$ alternate with planes of YZ along the 001 direction. Using ab-initio electronic structure calculations, we have investigated planar insertions along 001 into the $L2_1$ structure. For most scenarios, insertion of single or double atomic layers of Cr into $Co_2MnGe$ or $Co_2MnSi$ did not destroy the half-metallic property. One insertion of a Cr layer into $Co_2MnGe$ was observed to increase the gap. In fact, we observed that for a large number of insertions using various transition metals or combinations of transition metals and non-transition metals, the band gap in the minority channel at the Fermi energy remains. An ad hoc rule that seems to partially capture the tendency to form half-metals can be formulated as: "001 planar insertions that can plausibly yield 8 down spin electrons on the $X_2$ layer and 4 down spin electrons on the YZ layer yield half-metals".

A half-metal is a ferromagnetic material that is metallic in one spin channel, and insulating or semiconducting in the other spin channel. Since the interesting nature of these materials was first high-lighted by DeGroot in 1983[1], several classes of these materials have been discovered, typically by means of ab-initio electronic structure calculations. Among these are the so called half-Heusler alloys with $C1_b$ structure[2], the full Heusler alloys with $L2_1$ structure[3] and rutile structure $CrO_2$,[4] one of the few ferromagnetic (as opposed to ferrimagnetic) oxides. Proving experimentally that these systems are indeed half-metallic has been more difficult, often because the half-metallicity is a bulk property while the experimental probes involve surfaces or interfaces where the half-metallic feature is lost. Other difficulties arise due to imperfections in the stoichiometry and anti-site disorder. Recently, however, evidence of high spin polarization has accumulated particularly for the full Heusler alloys.[5,6,7,8,9]

Recently, it has been pointed out that both the half-Heusler and the full Heusler alloys satisfy a Slater-Pauling-like relation between the number of valence electrons per formula unit and the magnetic moment[10]. For the full Heuslers ($L2_1$ structure) the rule is that the magnetic moment per formula unit is given by M=Z-24. For the half-Heuslers (C1b structure) the rule is that the moment is given by M=Z-18. Since $Z=N^\uparrow+N^\downarrow$ and $M=N^\uparrow-N^\downarrow$, the rule for the full Heusler is equivalent to the statement that the number of occupied minority bands per formula unit is 12, and for the half-Heusler, 9.

In this paper, we consider the full Heusler alloys and elaborate on the observation that they have 12 occupied minority bands per formula unit below an energy gap. Actually, since one can obtain half-metallic Heusler alloys such as $Mn_2VAl$ which obey the M=Z-24 rule, but with M<0, the statement should be generalized slightly to state that there is a strong tendency for $L2_1$ alloys with structure $X_2YZ$ where X and Y are transition metal atoms and Z is a non-transition metal ligand, to have 12 occupied bands in one spin channel separated by an energy gap. The energy stabilizing the Heusler phases is likely due to the formation of this energy gap. Moreover, we show that substantial disruption and modification of the $L2_1$ structure, such as planar insertions or substitutions along the (001) direction preserve the minority energy gap leading thereby to a new class of half-metallic layered structures based on Heusler alloys. All of the calculations shown here employed Density Functional Theory within the Generalized Gradient Approximation as implemented in the VASP code.[11,12,13,14]

The $L2_1$ structure for $X_2YZ$ is usually described as an fcc bravais lattice with a four atom basis. Since we generally build our films layer by layer, it is helpful to think of the full Heuslers as bcc structures layered along the 001 direction (Fig. 1). The layers consist of square cells of 2 X atoms alternating with square cells containing Y and Z atoms. The elaboration on the rule of 12 minority states per formula unit is that the 12 states divide into approximately 8 per $X_2$ layer and approximately 4 per YZ layer. Thus, using notation shown in Fig.1 for $Co_2MnGe$ (X=Co, Y=Mn, Z=Ge) the number of up ($N^\uparrow$) and down ($N^\downarrow$) spin electrons per layer follows the pattern represented in the table in Fig. 2 where the corresponding density of states of $Co_2MnGe$ is also shown. The tendency to form this pattern is robust in the sense that given a reasonable opportunity to form the 8, 4, 8, 4 … pattern, the system will do so, and the driving energy for the pattern comes from the decrease in system energy due to the gap (Fig. 2). The total number of minority states below the Fermi energy is 12, in accordance with the Slater-Pauling rule. A gap of approximately 0.4 eV separates the occupied and empty minority states. We obtain similar results for other full Heusler alloys such as $Co_2MnSi$, $Co_2FeAl$, $Co_2CrSi$ etc., which have a very similar electronic structure. For the half-metallic full Heusler alloys the number of occupied minority states is 12. In order to test the robustness of the 8-4 rule, a modified cell of $Co_2MnGe$ having two atomic layers rather than four as in figure 1 was studied. This cell consists of a layer of Co alternating with a layer of MnGe. This cell is not an $L2_1$ structure, however it forms a half-metal with a gap of approximately 0.25 eV.

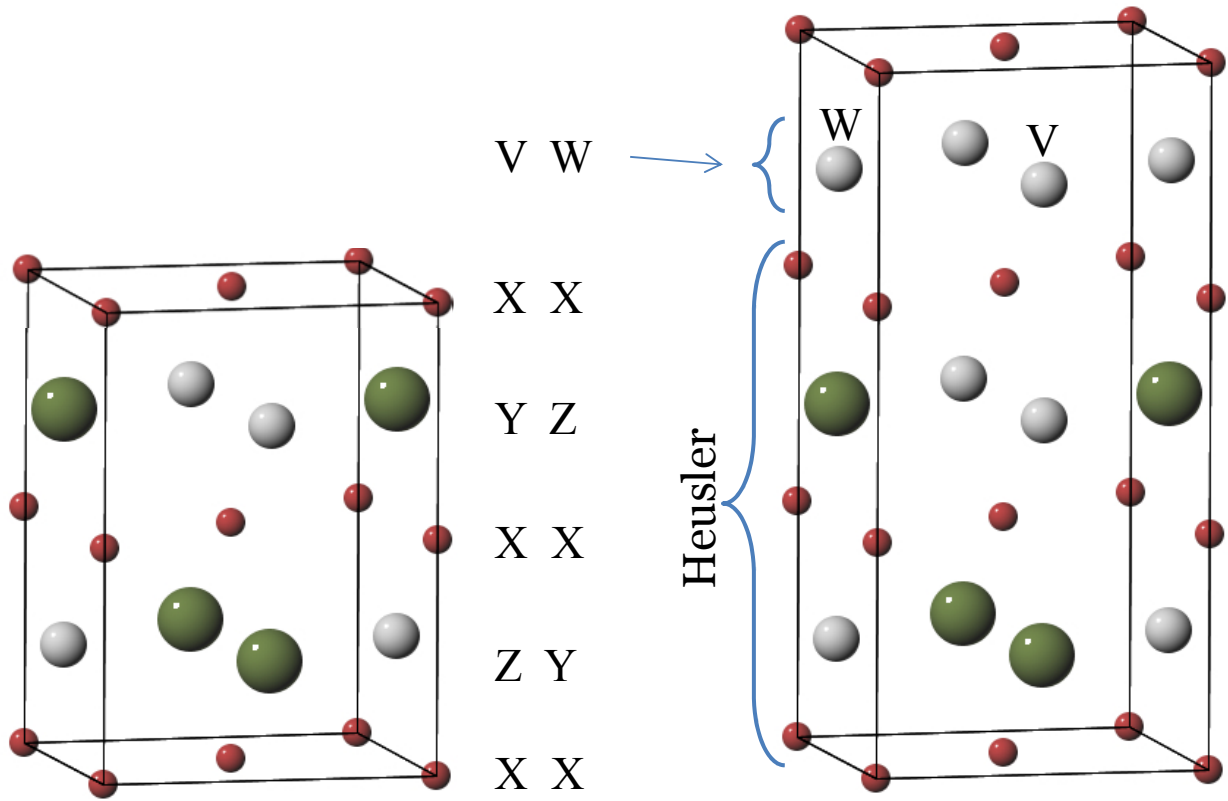

**Fig. 1. (Left)**: $L2_1$ structure with layers along (001) direction. Layers with two atoms of X per cell alternate with layers containing one atom each of Y and Z. The positions of Y and Z are interchanged on alternate layers. If Y and Z are the same atom, the structure becomes B2. If X, Y, and Z are the same atom, the structure becomes bcc. **(Right):** The same with inserted layer of atoms V and W.

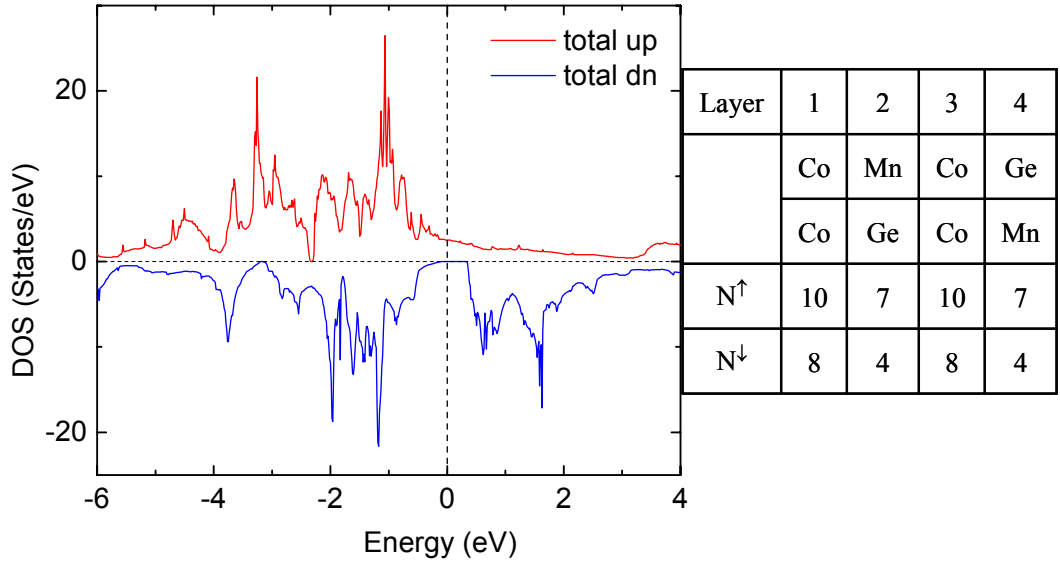

**Fig. 2.** Density of states of the full Heusler $Co_2MnGe$ with corresponding pattern of majority and minority electron distribution

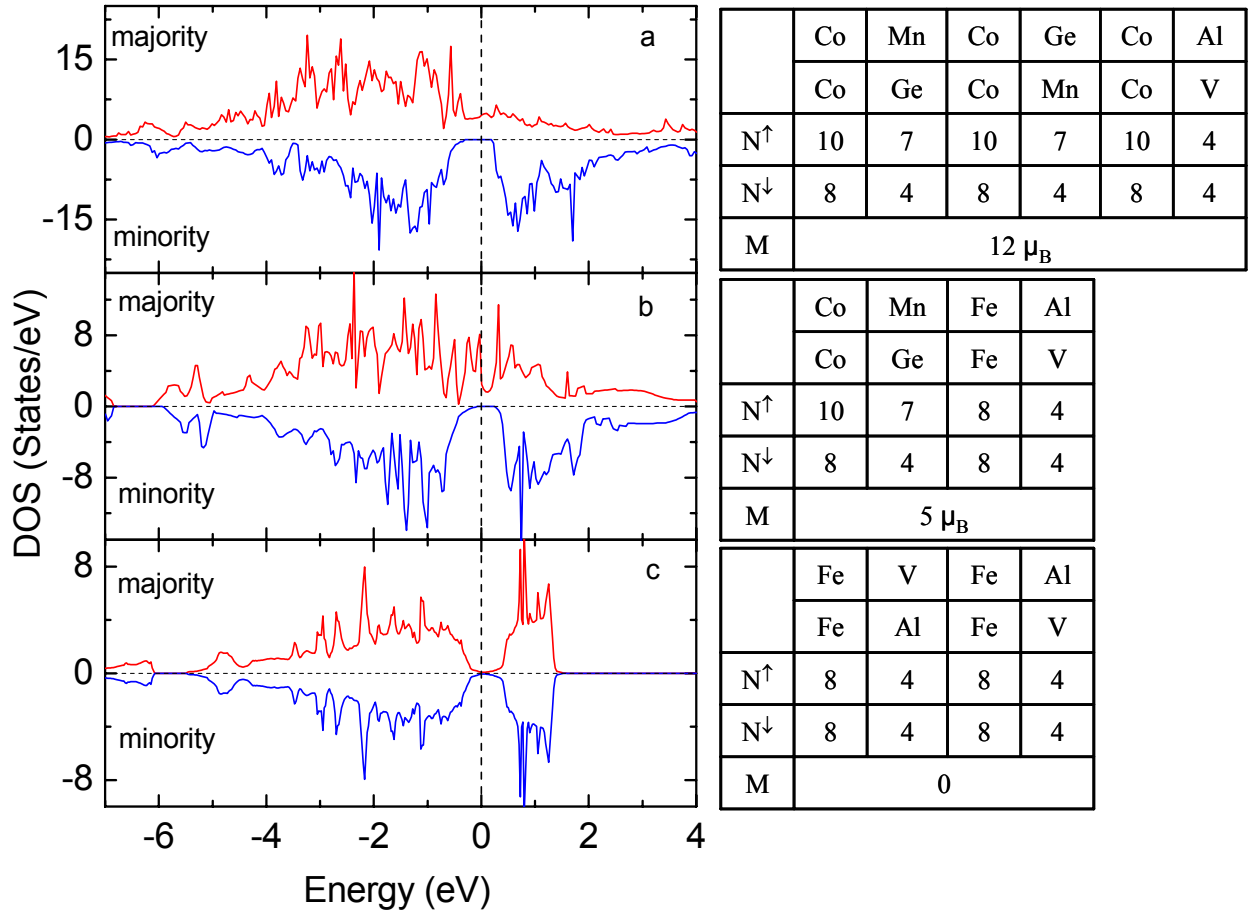

**Fig. 3.** Density of states for (a) 5 layers of $Co_2MnGe$ with one layer of VAl; (b) two layers of $Co_2MnGe$ with two layers of $Fe_2VAl$; (c) four layers of $Fe_2VAl$ full Heusler alloy.

In addition to dozens of known and previously proposed Heusler alloys, many more layered structures are also predicted to be half metallic. Some of the systems that we have verified (by calculation) to be half metallic with the calculated DOS and moments are shown in Fig. 3. In Fig.

3a we show density of states of a multilayer with 6 atomic layers per repeat. The first five layers $Co_2|MnGe|Co_2|GeMn|Co_2$, are a sequence from the $Co_2MnGe$, full Heusler. The sixth layer is VAl substituting for a MnGe layer. Note that the VAl layer is also half-metallic. It is able to generate 4 down spin electrons by splitting its 8 electrons equally between up and down spin. A second example is provided by a 4 layer repeating sequence, $Co_2|MnGe|Fe_2|AlV$. In this case, both the Fe layer and the AlV layer must give up their magnetic moment to generate the 8-4-8-4 sequence and retain the half-metal gap (Fig. 3b).

Some systems are not sufficiently polarizable to generate the required pattern. One example is $Co_2FeSi$ which has been proposed as a high moment half-metal[15]. As already observed, the Co atoms with 9 electrons (18 per layer) can polarize as 5up and 4down giving the required 8 down per layer (and 10 up). The FeSi layer, however with a total of 12 electrons can only generate 4 down spin electrons if the Fe polarizes as 6up and 2 down. This is very difficult since there are only 5 d-states. The remaining s-state is very difficult to polarize. Accordingly, our DFT-GGA calculations do not predict $Co_2FeSi$ to be a half-metal. Others have used LDA+U to generate a half-metal from $Co_2FeSi$[10].

The examples above are based on combining and mixing systems that would separately also be half-metals, e.g. $Co_2MnGe$ and $Co_2VAl$ are both separately predicted to be half-metals in the $L2_1$ phase, as are $Co_2MnGe$ and $Fe_2VAl$. Actually $Fe_2VAl$ doesn't quite survive as a half-metal as an infinite system (Fig. 3c). This is an interesting case because $Fe_2VAl$ is paramagnetic despite being 50 atomic percent Fe. This, of course is predicted by the 8,4 rule. The only way one get 8 and 4 in the down spin channel when the electron count in the two layers is 16 for the Fe layer and 8 for the VAl layer, is for the there to be no spin-polarization. Theoretically at the level of DFT-GGA (and apparently also experimentally[16,17,18]), the material is a semi-metal with only a tiny density of states at the Fermi energy. The maximum of the valence band at the center of the zone is only slightly higher than the minimum of the valence band at k = (0.5, 0.5, 0)π/a.

Even more surprising evidence of the 8-4 rule can be obtained using Cr interlayers, because 2 Cr atoms have a total of 12 electrons. This means they can generate both the 8 and the 4. Therefore they can substitute for a Co layer (8 minority electrons) or a MnGe layer (4 minority electrons). Figure 4b shows the density of states for the case in which Cr substitutes for Co in a 6 layer repeating sequence MnGe|Co$_2$|GeMn|Co$_2$|MnGe|Cr$_2$. Since the two Cr atoms substitute for

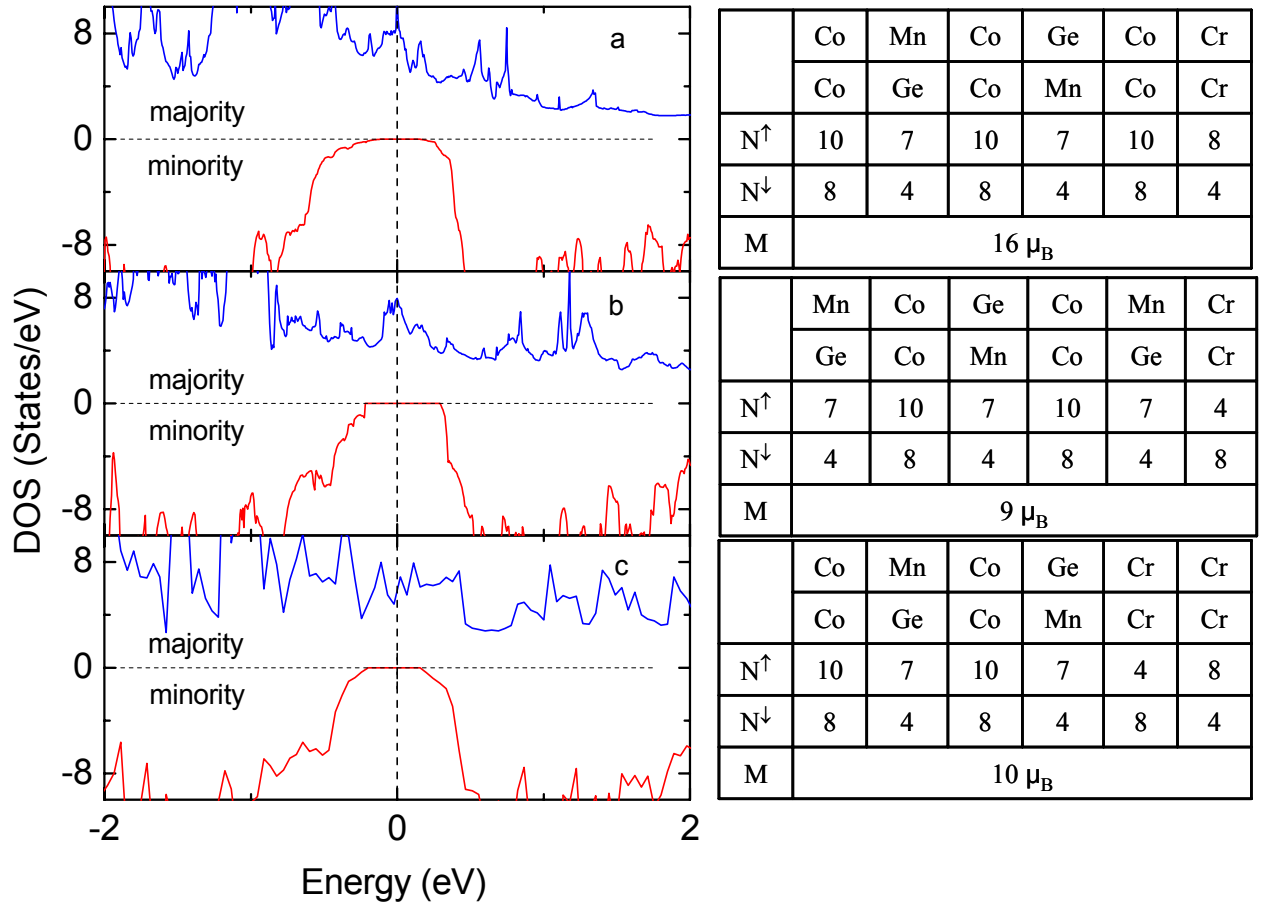

**Fig. 4.** Density of states of the full Heusler Co$_2$MnGe with planar one (a and b) and two (c) planar inserted Cr layers

Co, they must generate 8 down spin electrons to maintain the 8-4-8-4 repeat pattern. Since the Cr atoms have 6 electrons each, they polarize as 2↑, 4↓. This generates a moment opposite to that of the remainder of the system. Another interesting result is that the substitution of Cr, even though it violates the $L2_1$ structure and symmetries in the two ways described above, actually increases the energy gap from 0.4 to 0.5 eV. Figure 4a shows the calculated Density of states for a 6 atomic layer cell in which a Cr layer substitutes for a MnGe layer. In this case, the 2 Cr atoms must generate 4 down spin electrons leaving 8 up spin electrons. In this case the Cr moments are parallel to those of the remainder of the Heusler alloy. Finally, we generated a lattice in which 4 Heusler layers ($Co_2$|GeMn|$Co_2$|MnGe) alternate with 2 Cr atomic layers (Fig. 4c). The Heusler down spin pattern is 8-4-8-4. In order for the 2 Cr layers to continue this pattern, the first Cr layer must generate 8 down spin electrons and the second 4. This implies that the two moments of the two Cr layers are opposite. We have the strange situation in which (at least for 2 layers) the system is simultaneously half-metallic and anti-ferromagnetic.

It should be understood that the 8-4-8-4 pattern can only be defined approximately because the assignment of a particular amount of charge situated in a particular part of the cell to a particular atom is at best ambiguous, depending on how one decides to draw the boundaries between atoms. However, the rule seems to be approximately true using most "reasonable" means of dividing the space in the cell among its constituent atoms. For the systems discussed here, we assumed that all atoms in the cell had the same size.

This work was supported by NSF MRSEC Grant N$^o$ DMR 0213985, and by the INSIC EHDR Program.